\documentclass[fleqn,twoside]{article}
\usepackage{espcrc2}

\title{\bf Comments on ``Are Swift Gamma-Ray Bursts consistent with the Ghirlanda relation?",
by Campana et al.(astro--ph/0703676)  }

\author{G. Ghirlanda
\address[OAB]{Osservatorio Astronomico di Brera, via Bianchi 46, I--20387, Merate, Italy}, 
L. Nava\addressmark[OAB]
\address{Univ. degli Studi dell'Insubria, via Valleggio 11, I--22100, Como, Italy}, 
G. Ghisellini 
\addressmark[OAB], 
C. Firmani\addressmark[OAB]
\address{
Instituto de Astronom\'{\i}a, U.N.A.M., A.P. 70-264, 04510, M\'exico, D.F., M\'exico}}

\begin{document}

\begin{abstract}
\vskip 0.5 true cm
{\bf {\hskip 6 true cm ABSTRACT}} \\ 
\vskip 0.1 true cm
{\large In their recent paper, Campana et al. (2007)
found that 5 bursts, among those detected by $Swift$, are outliers 
with respect to the 
$E_{\rm peak}$--$E_{\rm \gamma}$ (``Ghirlanda") correlation.
We instead argue that they are not.}
\vspace{1pc}
\end{abstract}

\maketitle

\section{Introduction}

Campana et al. (2007, C07 hereafter) investigate the 
$E_{\rm peak}$--$E_{\rm \gamma}$ (so called
``Ghirlanda") correlation, including all GRBs
detected by $Swift$ for which we know
the redshift, the peak energy $E_{\rm peak}$ and
we have information on the presence of the jet break,
necessary to estimate the jet opening angle,
and therefore to calculate the collimation corrected
bolometric energy, $E_{\rm \gamma}$.
In a similar study performed by us
(Ghirlanda et al. 2007, G07 hereafter), we concluded
that there was no new outlier with respect to
the $E_{\rm peak}$--$E_{\rm \gamma}$ correlation 
(besides GRB 980425 and GRB 031203,
but see Ghisellini et al. 2006),
while C07 claim that there are 5 $Swift$ bursts which
do not obey the correlation. 
The sample of GRBs studied by C07 and G07 is the same.
In the following we give arguments contrasting the claim of C07.

\section{GRB 060526}

This burst is the second most important outlier (in term of contribution to
the $\chi^2$) presented by C07.
Both C07 and G07 used the same source of data: Schaefer (2007)
for the fluence and $E_{\rm peak}$, and Dai et al. (2006) for $t_{\rm jet}$.
Using the listed bolometric fluence one obtains
$E_{\rm \gamma, iso}=2.53\times 10^{52}$ erg.
We recomputed the bolometric fluence from the spectral
parameters reported by Schaefer (2007), obtaining  
$E_{\rm \gamma, iso}=2.58\times 10^{52}$ erg, which is the value we used.
Instead C07 list an
isotropic energy $E_{\rm \gamma, iso}= 1.07^{+0.16}_{-0.14}\times 10^{53}$ erg.
We remind that the isotropic energy is found through
\begin{equation}
E_{\rm \gamma, iso} \, =\, S_{\rm bol} \, {4\pi d_{\rm L}^2 \over (1+z)}
\end{equation}
where $S_{\rm bol}$ is the bolometric fluence and the $(1+z)$ term accounts
for the cosmological time dilation. {\it Neglecting the $(1+z)$ term,} and
using the bolometric fluence $S_{\rm bol}=1.17\times 10^{-7}$, as listed by
Schaefer (2007), one obtains $E_{\rm \gamma, iso}=1.07 \times 10^{53}$ erg,
which is the value reported in C07.  We therefore suggest that C07, for this
burst, missed the $(1+z)=4.21$ term when calculating $E_{\rm \gamma, iso}$.
The $E_{\gamma}$ value used by C07 is therefore larger than the value
found by G07 because of the larger $E_{\rm \gamma, iso}$ ($t_{\rm jet}$ is the same).

A separate problem concerns the  values of $E_{\rm peak}$ and
bolometric fluence for this burst reported by Schaefer (2007).
In fact, this burst showed two main peaks in BAT, separated by $\sim$200 seconds,
with the second peak having a slightly larger fluence than the first,
with a softer spectrum.
The spectral behaviour of this burst is thus complex, and the value of $E_{\rm peak}=25\pm 5$
keV reported by Schaefer (with a fluence corresponding to the first peak only)
may be controversial.
For this reason we have analyzed the available $Swift$ data for this burst.
Our results and the consequences for the Ghirlanda correlation  can be found at:
www.brera.inaf/utenti/gabriele/060526/060526.html .

\section{GRB 050922C and GRB 060206}

These two bursts lack optical data at times late enough 
to encompass the jet break time predicted by the Ghirlanda relation.
The fact that there is indeed an {\it early} break in the optical
{\it does not guarantee} that this is a jet break, since
we now know that there is the possibility of multiple breaks 
in the optical.
In these cases only a lower limit on the break time can be taken,
corresponding to the latest optical observations, as discussed
in G07.

\section{GRB 050401 and GRB 050416A}

Several authors published a partial coverage of the optical
afterglow of these two bursts, but none of them discussed the
results which can be obtained by collecting all the available
data (at least in one band).
Therefore, the claim that in these GRBs there is no apparent break
refers to the partial coverage presented in each paper.
Because of that, in G07 we constructed the light curves with data from 
different sources.

In GRB 050401 the result of the fitting is somewhat dependent
from the (yet unknown) assumed magnitude of the host galaxy,
which can contribute to the late photometric points.
Furthermore, there is a large uncertainty in the normalisation
of the De Pasquale et al. (2006) points, because they used a different
reference star for their differential photometry.
What we plotted in Fig. 1 of G07 assumes the maximum possible 
displacement (--0.5 mag): assuming a lower one would inevitably
worsen a single power law decay fit.

For GRB 050416A, it is true that Soderberg et al. (2006) stated
that a single power law decay plus a 1998bw--like supernova
light curve can fit the data, but also in this paper there is
no complete collection of points coming from the available literature.
Anyway, SN 2006aj associated with GRB 060218 is by far the best studied
at early times, so using this as a template should give a more reliable
result. In this case the presence of a break in the optical light curve
is clearly required.

Given all the above, we think that
in these two GRBs there exists a margin of
subjectivity for judging the presence or not of a possible jet break
(this margin is however small for GRB 050416A).
But just because of this, it is not appropriate to declare that
they are outliers, and treat them as such in the fits.
At the very least, one should consider them having a lower limit in $E_{\gamma}$
corresponding to the jet break time we have derived.

\section{Additional comments}

The pre--Swift data plotted in the figures of C07 are 
the values of $E_{\gamma}$ calculated taking 
$E_{\rm \gamma, iso}$ from Firmani et al. (2006) and
the jet angles from Nava et al. (2006), who reported
slightly different values of $E_{\rm \gamma, iso}$.
Since the derived jet angle depends upon $E_{\rm \gamma, iso}$,
this procedure is not correct.

When calculating the $\chi^2_r$ value for the bursts in 
the sample of Nava et al. (2006), C07 find agreement in the
case of an homogeneous circumburst medium, and a larger
$\chi^2_r$ in the case of a wind profile.
We instead confirm the original value reported
in Nava et al. (2006).

We  note that the $\chi^2_r$ values given in 
Table 2 of C07 for the ``$Swift$ data achromatic breaks"
and ``$Swift$ data pure breaks" cases, do not correspond to the values
given in the text.

GRB 061121 is plotted as a lower limit in $E_{\gamma}$, and lies
to the left of the Ghirlanda correlation. It should not be included
in the fit as instead done in C07.

A symmetric error on a linear quantity transforms into
an asymmetric error in the logarithm.
We believe that C07 underestimated the error on $E_{\rm \gamma, iso}$
due to the systematic choice of the smallest error in the logarithmic quantity.
In G07, instead, we propagated the errors in the logarithmic space.

Finally, in Fig. 2 of C07 (wind case) there is an additional
pre--$Swift$ burst which is not present in Fig. 1.

\section{Conclusions}

We would like to stress that we are not willing to defend the 
$E_{\rm peak}$--$E_{\gamma}$ correlation to death.
As any other scientific result, it must be the object of severe
scrutiny from the scientific community.
This is even more true since its potential
cosmological use makes this correlation very important 
[as well as the related, model independent 
and assumption free, Liang \& Zhang (2005) correlation].
Furthermore, its existence can flag some crucial property
of GRB physics which are not yet fully understood  
(but some attempts have already been done, see Thompson 2006
and Thompson, Meszaros \& Rees 2007).
Therefore to demonstrate that this correlation is the result
of some selection effects (or not), or that its dispersion
is much larger than what it is now (or not), or that there are
outliers (or not) is a {\it mandatory} task, that must be
pursued {\it carefully}.

\end{document}